\documentclass[a4paper]{article}
\pdfoutput=1
\usepackage[affil-it]{authblk}

\usepackage[latin1]{inputenc}
\usepackage[T1]{fontenc}
\usepackage[english]{babel}
\usepackage{amsfonts}
\usepackage{amssymb}
\usepackage{amsmath}
\usepackage{amsthm}
\usepackage{graphicx,xcolor}
\usepackage{a4wide}

\usepackage{esint}

\makeatletter

\@ifundefined{textcolor}{}
{
 \definecolor{BLACK}{gray}{0}
 \definecolor{WHITE}{gray}{1}
 \definecolor{RED}{rgb}{1,0,0}
 \definecolor{GREEN}{rgb}{0,1,0}
 \definecolor{BLUE}{rgb}{0,0,1}
 \definecolor{CYAN}{cmyk}{1,0,0,0}
 \definecolor{MAGENTA}{cmyk}{0,1,0,0}
 \definecolor{YELLOW}{cmyk}{0,0,1,0}
}

\usepackage{bbm}
\usepackage{psfrag}
\usepackage{color}

\makeatother

\begin{document}

\title{Centrality metrics and localization in core-periphery networks}

\author{Paolo Barucca}
 \affil{%
 Scuola Normale Superiore, Italy  }%
\author{Daniele Tantari}%
\affil{%
Centro di Ricerca Matematica Ennio De Giorgi, \\
Scuola Normale Superiore, Italy}%
\author{Fabrizio Lillo}%
\affil{%
Scuola Normale Superiore, Italy}%

\maketitle

\begin{abstract}

Two concepts of centrality have been defined in complex networks. The first considers the centrality of a node and many different metrics for it has been defined (e.g. eigenvector centrality, PageRank, non-backtracking centrality, etc). The second is related to a large scale organization of the network, the core-periphery structure, composed by a dense core plus an outlying and loosely-connected periphery. In this paper we investigate the relation between these two concepts. We consider networks generated via the Stochastic Block Model, or its degree corrected version, with a core-periphery structure and we investigate the centrality properties of the core nodes  and the ability of several centrality metrics to identify them. 
We find that the three measures with the best performance are marginals obtained with belief propagation, PageRank, and degree centrality, while non-backtracking and eigenvector centrality (or MINRES \cite{key-31}, showed to be equivalent to the latter in the large network limit) perform worse in the investigated networks.
\end{abstract}


\section{Introduction}

Network theory now represents a standard framework to describe and analyze complex systems, from biology to finance \cite{watts}. In complex systems single elements can be influenced both by closest neighbors (microscopic), by the behavior of the group they belong to (mesoscopic), and by the properties of the network as a whole (macroscopic). Thus quantifying properties at different scales is fundamental to understand the underlying  systems and hopefully provide predictive models for complex phenomena. 

In particular, centrality is a pivotal topic in network theory because it encompasses all different scales in networks: central nodes often play important roles in systems' growth \cite{BARABASI,anergy}, cascade effects \cite{Buldyrev}, information and epidemic spreading \cite{pastor}. Moreover given the large amount of Big Data nowaday accessible for scientific research in a broad range of fields, identifying in a fast and statistically robust way a subset of relevant, i.e. central,  elements is crucial for being able to carry out detailed and insightful analysis. 
Nevertheless the notion of relevance can be different in different systems and for this reason various measures of centrality have been introduced in the past years: degree centrality, eigenvector centrality, PageRank \cite{PR1}, non-backtracking centrality \cite{newman2}, etc. One critical aspect of node centrality measures is the role of localization in presence of hubs. This means that, in the presence of strong degree heterogeneity, only few components of most centrality measures are significantly different from zero. This last property has been conjectured to influence the sensitivity of a given centrality measure  \cite{newman2} and some metrics, such as non-backtracking centrality, have been recently introduced to mitigate this problem.

While these centrality measures consider individual nodes, core-periphery is a mesoscopic notion of centrality. A core-periphery network has a block structure made by an high link-density core and an outlying low-density periphery. Core-periphery structure is relevant because it represents one of the possible large scale organizations of networks different from the modular organization in community.   Core-periphery structures have been documented in a variety of systems described by networks. Originally introduced in sociology \cite{key-29,key-30,key-31} they have also been identified in international trade \cite{smith}, and in finance, especially in the interbank network \cite{FrickeLux,lelyveld} (but see \cite{barucca} for a recent critical analysis).  Various methods have been proposed to identify the two blocks and to quantify coreness, i.e. the probability that a node belongs to the core. Among the methods we mention MINRES \cite{key-31} and stochastic block model (SBM) probability marginals \cite{newman1}.

The two concepts of centrality described above are conceptually different but clearly related one with the other. The purpose of this paper is to investigate the relation between node centrality measures and core periphery structure. Since this relation is generically dependent on the investigated system, here we study the node centrality properties of a model network with a core periphery structure. More specifically, we consider networks generated according to the SBM and with a clear core-periphery structure and we investigate (i) the node centrality properties of core nodes and (ii) how well different node centrality metrics are able to identify the core nodes. The considered metrics are probability marginals of SBM obtained with Belief Propagation, Degree centrality, Eigenvector centrality, MINRES, Non-Backtracking centrality, and PageRank. We investigate also the role of localization in the performance of these methods. Unlike previous works we analyze also the case of degree-heterogeneous core-periphery networks sampled from a degree-corrected SBM. We find a clear ranking of the performance of the different methods and distinct node centrality properties of the core nodes. 

More specifically, in Section \ref{sec:method} we introduce the generative model and the details of the centrality measures, discussing their mutual relations and expected differences. In Section \ref{sec:hom} we consider an ensemble of core-periphery SBM networks, without degree-correction, and verify the differences between centrality measures in this case. In particular for the first time, at best of our knowledge, we observe analytically and numerically that eigenvector centrality and MINRES coreness coincide in the infinite-size limit. In Section \ref{sec:het} we consider a degree-corrected SBM and we compare the performance of the different centrality measure. As a major result we find that PageRank surpasses SBM marginals in strongly degree heterogeneous graphs. As a real case study, we compare different centralities on a
network constructed from data from the Oregon Routeviews Project \cite{oregon} and verify the distribution of their elements and their mutual correlation. In Section \ref{sec:concl}  we summarize our results and indicate various open questions related to  the properties of centrality measures in large directed heterogeneous networks.

\section{Methods}\label{sec:method}

\subsection{The generative model}
Core periphery structure in a network is the presence of a significant division of the nodes in two groups, one relatively densely connected group, the core, and one less densely connected group, the periphery, mostly connected with the core. 
The most common way to define such kind of structures with a rigorous but at the same time flexible characterization is to introduce a stochastic generative model, i.e. a model able to create artificial networks with certain properties that can be fitted to real network data. In particular, core-periphery can be seen as special case of a community based structure and for this purpose the Stochastic Block Model (SBM) \cite{SBM1,SBM2} is the most established and widely used generative model.

In an SBM of $N$ nodes, one creates different communities labelling each of them (for the aim of this work we just consider the case of two communities representing respectively the core and the periphery). Then each node is assigned randomly to the group $1$ (the core) with a probability $\gamma_1=\gamma$ or to the group $2$ (the periphery) with probability $\gamma_2=1-\gamma$. Finally, for each couple of nodes $(i,j)$ owning to the groups $(g_i,g_j)$, with $g_,g_j \in \{1,2\}$, we add an undirected link independently with probability $p_{g_ig_j}$ according to the entries of the so called \textit{affinity matrix} $p_{ab}$. Since the graph is undirected, the affinity matrix $p$ is a $2\times 2$ symmetric matrix and consequently the model has four independent parameters: the expected size of the core $\gamma$ and the expected degree between nodes belonging to different communities $p_{11}$, $p_{12}$ and $p_{22}$.  Thus, given $\gamma$ and $	p$, we are defining an ensemble of graphs according to
\begin{equation}\label{eq:ML}
\mathcal{P}(A,\boldsymbol{g}|p,\gamma)=\prod_{i=1}^N \gamma_{g_i} \prod_{(i,j)}^N p_{g_ig_j}^{A_{ij}}(1-p_{g_ig_j})^{1-A_{ij}},
\end{equation}
where $A_{ij}$ is the adjacency matrix of the network having value one if there is an
edge between nodes $i$ and $j$ and zero otherwise. Since we are interested in sparse graph we rescale tha affinity matrix to be $N p_{ab}= c_{ab} $, such that the mean degree is finite $\langle d\rangle=\sum_{a,b={1,2}}\gamma_a\gamma_{b}c_{ab}$.

Depending on the ranking between the elements of the affinity matrix, the network has different structure. We refer to core periphery structure when  $p_{11}>p_{12}>p_{22}$, i.e. when links are most probable in the core and nodes in the periphery are more likely connected to those in the core than to each other. This last property distinguishes a core-periphery structure from assortative ($p_{11}>p_{12}<p_{22}$) and dissortative ($p_{11}<p_{12}>p_{22}$) ones.  Note these relations involve the $p$'s entries, i.e. only for the mean degrees. Less flexible generative models can be obtained  by requiring that every core-core, core-periphery, and periphery-periphery degree satisfy the previous order relation, for example constraining them at fixed ordered values (Regular Stochastic Block Model (r-SBM))\cite{SBM2}.

One of the limits of SBM is that degree distribution cannot be fat tailed because it is essentially a superposition of Poisson distribution. In many real systems core periphery structure is present also in the presence of an highly heterogeneous core, i.e. a densely connected group of nodes with further highly connected hubs.  We introduce a sequence of numbers (degree corrections) $\boldsymbol{w}=(w_1,\ldots, w_n)$ from a given probability distribution and associate each of them to a node. Then we consider random graphs in which edges are independently assigned to each pair of nodes $(i,j)$ with probability $w_iw_j\rho \ c_{g_ig_j}$ where $\rho=(\sum_j w_j)^{-1}$. In this way the expected core-core degree (given the corrections) of the $i$-th node is $c_{11}\gamma w_i$ and the degree distribution is determined by the one from which the degree corrections are drawn. This generative model is the Degree Corrected Stochastic Block Model (dc-SBM) \cite{SBM1}. It reduces to the standard SBM if the corrections are set all $1$, while it allows to generate networks with heterogeneous core-periphery structure if the $w_i$ are drawn from a broader probability distribution. In Section \ref{sec:het} we use a power law distribution with tail exponent $\alpha>2$ and unitary mean by choosing the degree sequence $\boldsymbol{w}$ satisfying $w_i=ci^{-1/(\alpha-1)}$ for $i=i_0,\ldots,i_0+N$ \cite{chung,strogatz}. Moreover, since we want the hubs to appear only in the core, we assign the nodes with the largest degree corrections to the core, ensuring that nodes with largest expected degrees stay there.  

\subsection{Measures of centrality}

In this Section we present the investigated centrality measures, namely the probability marginals of SBM obtained with Belief Propagation (BP), Degree centrality (DEGREE), Eigenvector centrality (EC), MINRES, Non-Backtracking centrality (NBT), and PageRank (PR).  
 
\subsubsection{Inference and belief propagation}

One of the advantages of SBM is that it is possible to infer the parameters by using Maximum-Likelihood methods \cite{key-35}.
The probability, or likelihood, that the network
was generated by the model is, given uniform priors, proportional to
\begin{equation}
\mathcal{P}(A|p,\gamma)=\sum_{\boldsymbol{g}}\mathcal{P}(A,\boldsymbol{g}|p,\gamma)\label{SBM-1}
\end{equation}
with $\mathcal{P}(A,\boldsymbol{g}|p,\gamma)$ as in Eq. ($\ref{eq:ML}$).
Maximizing the loglikelihood  with respect to
$p_{ab}$ and $\gamma_a$ we obtain the most likely values of the parameters given by
\begin{align}
p_{ab} & =\frac{\sum_{ij}A_{ij}q_{ab}^{ij}}{\sum_{ij}q_{ab}^{ij}},\label{eq:prs0}\\
\gamma_{a} & =\frac{1}{N}\sum_{i}q_{a}^{i},\label{eq:gamma}
\end{align}
where we have defined the, $(p,\gamma)$ dependent,  one node and two nodes marginal probabilities
\begin{align*}
q_{a}^{i} & =P(g_{i}=a)\\
q_{ab}^{ij} & =P(g_{i}=a,\ g_{j}=b|A_{ij}=1)
\end{align*}
computed starting from the assignment probability distribution
\begin{equation}
P(\boldsymbol{g}|p,\gamma,A)=Z^{-1}
\prod_{i=1}^N \gamma_{g_i} \prod_{(i,j)}^N p_{g_ig_j}^{A_{ij}}(1-p_{g_ig_j})^{1-A_{ij}},
\end{equation}
where $Z=\mathcal{P}(A|p,\gamma)$ is a normalization constant. Solving equations $(\ref{eq:prs0},\ref{eq:gamma})$ gives at the same time the optimal fit parameters (including the core size $\gamma$) and the assignment probability distribution with its one node marginals $q_{a}^{i}$. In particular $q_{1}^{i}$ can be interpreted as the coreness of the node $i$ and ordering nodes with decreasing  coreness allows to detect which nodes are most likely to be in the core, consistently with the estimated value of $\gamma$.

Equations $(\ref{eq:prs0},\ref{eq:gamma})$ are usually solved iteratively as follows: 1) make an initial guess for the fit parameters $(p,\gamma)$; 2) compute the assignment probability distribution and the marginals given $(p,\gamma)$; 3) use equations $(\ref{eq:prs0},\ref{eq:gamma})$ to calculate an improved estimate of  $(p,\gamma)$ and repeat from step 2 until convergence.

Step 2 is the most critical one because it is not possible to compute $Z$ directly by evaluating the sum over all the possible assignments. In fact this sum has an exponentially large (in $N$) number of terms and it
would take prohibitively long time to compute it numerically.  To overcome this problem one can use a Monte Carlo sampling \cite{MC1} or can implement an approximated method for the
marginal probabilities. The most common is the one introduced by Decelle at al. \cite{SBM2,SBM3} based on a message passing alghoritm or \textit{belief propagation} (BP). 
The idea is to define the cavity marginals, or messages, $\eta_{a}^{i\rightarrow j}$ which is equal to the probability that node $i$ belongs to group $a$ once node $j$ is removed from the network. Removing
$j$ allows to derive a set of self-consistent equations that must be satisfied
by these messages, reading as \cite{SBM2}
\begin{equation}
\eta_{a}^{i\rightarrow j}=\frac{1}{\mathcal{Z}^{i\rightarrow j}}\gamma_{a}\prod_{\substack{k\sim i\\
k\ne j
}
}\sum_{b}\eta_{b}^{k\rightarrow i}p_{ab}\prod_{\substack{k\not\sim i\\
k\ne j
}
}\sum_{b}\eta_{b}^{k\rightarrow i}(1-p_{ab})\label{eq:bp-update},
\end{equation}
where $i\sim j$ stands for $A_{ij}=1$ and $\mathcal{Z}^{i\rightarrow j}$ is a normalization constant ensuring $\sum_a \eta_{a}^{i\rightarrow j}=1$ . As usual in BP, this expression is based on a locally tree like approximation of the graph, i.e. assumes that nodes other than $i$ are independent conditioned on $g_{i}$. This means that BP works better for a sparse graph, where loops are unlikely.

Once the belief propagation equations have converged, the BP estimates of the marginals can be expressed in terms of the messages as
\begin{equation}
q_{ab}^{ij}=\frac{1}{\mathcal{Z}^{ij}}\eta_{a}^{i\rightarrow j}\eta_{b}^{j\rightarrow i}\left\{ \begin{array}{ll}
p_{ab} & \mbox{if }i\sim j\\
1-p_{ab} & \mbox{otherwise}
\end{array}\right.\label{eq:marginals}
\end{equation}
\begin{equation}
q_{a}^{i}=\frac{1}{\mathcal{Z}^{i}}\gamma_{a}\prod_{\substack{k\sim i}
}\sum_{b}\eta_{b}^{k\rightarrow i}p_{ab}\prod_{\substack{k\not\sim i}
}\sum_{b}\eta_{b}^{k\rightarrow i}(1-p_{ab})\label{eq:marginals2},
\end{equation}
where again $\mathcal{Z}^{i}$ and $\mathcal{Z}^{ij}$ are constants ensuring  normalization for $q_{a}^{i}$ and $q_{ab}^{ij}$.

\subsubsection{Degree centrality}
The advantage of the Maximum-Likehood approach is that it gives as output both the size of the core $\gamma$ and the marginals $q^i_a$, interpreted as a measure of the coreness of the node $i$.
Now suppose we already know which is the right size of the core. Then we would need just a measure of the coreness (or centrality) of each node to choose the most likely assignment. The simplest measure of centrality we can think is based exclusively on the degree. We can order the nodes by decreasing degrees,
$$
k_1\geq k_2\geq\ldots \geq k_N,
$$  
and assign the first $\gamma N$ of them to the core. Albeit simple, this kind of centrality captures very well the coreness of a node in a SBM network, since, in some particular regimes, BP marginals centrality reduces to it. In fact, as shown in \cite{key-35}, defining the odd ratio $q^i_1/q^i_2$ between the probabilities of node $i$ to be in the core or not, expanding the last product of  equation $(\ref{eq:marginals2})$ in the first order in $p_{ab}=c_{ab}/N$ and using $\eta_{a}^{k\rightarrow i}=q^i_a + o(1/N)$ if $k\not\sim i$, we get
\begin{equation}
\frac{q_{1}^{i}}{q_{2}^{i}}=\frac{\gamma_{1}}{\gamma_{2}}e^{-k_{1}+k_{2}}\underset{k\sim i}{\prod}\frac{\eta_{1}^{k\rightarrow i}c_{11}+\eta_{2}^{k\rightarrow i}c_{12}}{\eta_{1}^{k\rightarrow i}c_{21}+\eta_{2}^{k\rightarrow i}c_{22}}\label{eq:bp-update-1},
\end{equation}
where $k_{1}$ and $k_{2}$ are respectively the averaged degree in the core and in the periphery, i.e. $k_a=\sum_b c_{ab}\gamma_b=1/N\sum_b c_{ab}\sum_i q^i_b$. Thus both when the structure is very weak ($c_{11}\sim c_{12}\sim c_{22}$) or robust ($c_{11} \gg c_{12}\gg c_{22}$), or in the particular case in which $c_{11}/c_{12}=c_{12}/c_{22}$, the ratio depends  only on the degree of the node $i$. This means that nodes with high degree are in the core while those with low degree are confined to the periphery. 

\subsubsection{Eigenvector centrality}

As we have shown, there is a clear correlation between degree and coreness of a node,
yet not an identity. What is missing is
the interconnectivity between core members. In other words, we should look at the core degree $k_{i}^{c}=\underset{g_j=1}{\sum}A_{ij}$, rather than at the  global degree, but we cannot evaluate the core degree without knowing the core. Therefore we might want to look at other centrality metrics. 

The most widely used of all this kind of recursive quantities, in which the coreness of a node is a function of the other nodes coreness itself,  is the \textit{Eigenvector centrality} \cite{EC1,EC2}, a centrality score $u_i$ proportional to the sum of the scores of the node's neighbors $\sum_j A_{ij} u_j$, i.e. a centrality vector $\boldsymbol{u}$ which is an eigenvector of the adjacency matrix
\begin{equation}
\sum_{j=1}^NA_{ij}u_{j}=\lambda_{1}u_{i}\label{EC}.
\end{equation}
associated to the largest eigenvalue $\lambda_{1}$. By the  Perron-Frobenius theorem \cite{PF} the corresponding eigenvector $\boldsymbol{u}$ is non negative. For an unweighted graph we can write ($\ref{EC}$) as
\begin{equation}
\underset{j\in N(i)}{\sum}u_{j}=\lambda_{1}u_{i}\label{EC-1}
\end{equation}
where $N(i)$ is the set of the neighbors of $i$. If we imagine to solve $(\ref{EC-1})$ recursively starting from an uniform solution at step zero $u_{i}^{(0)}=const$.
Then at step one we have: $k_{i}u^{(0)}=\lambda_{1}u_{i}^{(1)}$, so that
each component has increased proportionally to its degree, while at step
two the effect will re-weight the contribution of neighbors by their
degree as well and so on. Note that the first eigenvector can also be detected through a variational principle as follow
\begin{eqnarray}\label{eq:ec1}
\boldsymbol{u}^*&=&\underset{\|\boldsymbol{u}\|=1}{\operatorname{argmax}} \left( \sum_{i,j=1}^N A_{ij}u_iu_j\right )\nonumber \\
&=&\underset{\boldsymbol{u}}{\operatorname{argmax}}\left(\sum_{i,j=1}^NA_{ij}u_iu_j-\lambda\sum_iu_i^2\right),
\end{eqnarray}
with $\lambda$ acting as a Lagrangian multiplier. The maximization of the Lagrangian gives the secular equation $(\ref{EC})$. 

\subsubsection{MINRES centrality}
Equation $(\ref{eq:ec1})$ seeks $\boldsymbol{uu^{T}}$, by maximizing its overlap $f(A,B)=\sum_{ij}A_{ij}B_{ij}$ with the adjacency matrix of the data, where the pattern matrix $\boldsymbol{uu^{T}}$ has large values for pairs of nodes that are both high in coreness, middling values for pairs of nodes in which one is high in coreness and the other is not, and low values for pairs of nodes that are both peripheral. It is the continuous version of the discrete matrix $\delta_{ij}$ equals $1$ if $i$ and $j$ are in the core and $0$ otherwise, used by Borgatti and Everett in \cite{key-29,key-30} to identify core-periphery structure.  $\boldsymbol{uu^{T}}$ also approximates the adjacency matrix by minimizing the sum of its residuals (MINRES) $ \sum_{ij}(A_{ij}-u_iu_j)^2$. Boyd et. al. \cite{key-31} proposed an alternative formulation of coreness by considering only the off-diagonal residuals:
\begin{equation}
\boldsymbol{u}^*=\underset{\boldsymbol{u}}{\operatorname{argmin}}\ H_{B}[\boldsymbol{u}]=\underset{\boldsymbol{u}}{\operatorname{argmin}}\sum_{i\neq j}(A_{ij}-u_{i}u_{j})^{2}\label{BOYD}
\end{equation}
If we introduce a coefficient $\lambda$ (irrelevant for the minimization) in front of $u_{i}u_{j}$ and expand $(\ref{BOYD})$ we obtain
\begin{equation}
H_{B}[\boldsymbol{u}]=K-\sum_{i,j=1}^N 2\lambda A_{ij}u_{i}u_{j}+\lambda^{2}(\sum_{i=1}^Nu_{i}^{2})^{2}-\lambda^{2}\sum_{i=1}^N u_{i}^{4}\label{BOYD-1}.
\end{equation}
where $K$ is the number of links.
Comparing with equation $(\ref{eq:ec1})$, in ($\ref{BOYD-1}$) there is an additional term $\sum_i u_i^4$ beyond the usual term depending on the $L^2$ norm of $\boldsymbol{u}$. It is called the \textit{inverse participation ratio} of the vector $\boldsymbol{u}$ and is a measure of its localization (see Eq. \ref{IPRdef} below for the definition).
Looking for the stationary points, we have an equation that is close to the secular equation
for eigenvalues with an additive cubic term
\begin{equation}
\sum_{j=1}^N A_{ij}u_{j}=\lambda u_{i}(\sum_{i}u_{i}^{2})-\lambda u_{i}^{3}\label{derBOYD-1}
\end{equation}
Thus, following Boyd, we look for solutions close to the first eigenvector
of $A$. As we are going to see, this difference is negligible if $A$ has a non-localized eigenvector, while it becomes more evident as soon as the occurrence of heterogeneity in the core degrees increases localization.  

\subsubsection{Non-Backtracking centrality}
In Ref. \cite{newman2}, following \cite{NBT1,redemption}, Newman {\it et al.} proposed an important change to the standard eigenvector centrality. 
They define the centrality of node $j$ as the sum of the centralities of
its neighbors calculated in the absence of node $j$. The main idea is to neglect the reflection mechanism bringing localization on hub, for which highly connected node
gives high centrality to its neighbors, which in turn, reflecting it back, inflate the hub's centrality.
This kind of centrality can be measured as the first eigenvector of the Non Back Tracking (NBT) operator $B$ introduced by decomposing $A_{ij}^{2}$ in its diagonal and off-diagonal parts, namely:
\begin{equation}
A_{ij}^{2}=k_{i}\delta_{ij}+\sum_{k}(1-\delta_{ij})A_{ik}A_{kj}=k_{i}\delta_{ij}+\sum_{k}B_{k\rightarrow j,i\rightarrow k}\label{nonback}
\end{equation}
More explicitly, given a list of $K$ edges, the non-backtracking operator between directed
edges is defined as the $2K\times2K$ matrix
\begin{equation}
B_{k\to l, i\to j}=\delta_{kj}(1-\delta_{il})\label{nonback-1}
\end{equation}
The element $v_{i\to j}$ of the first eigenvector of $B$ gives  the centrality of the node $i$ in the absence of signals from  $j$. The full non-backtracking centrality $u_j$ of the node $j$ is defined as $u_j=\sum_i A_{ij}v_{i\to j}$. One can directly compute $\boldsymbol{u}$, without passing through the diagonalization of $B$, as the first $N$ entries of the leading eigenvector of the $2N\times 2N$ matrix
\begin{equation}
\left(\begin{array}{cc} 
A&1-D\\
1&0
\end{array}\right),
\end{equation}
where $A$ is the adjacency matrix and $D$ is the diagonal matrix with the degree of the nodes along the diagonal.

\subsubsection{PageRank}
Another centrality measure is PageRank, an algorithm used by Google Search  \cite{PR1} to rank website pages in their search engine results. The numerical weight it assigns to any given element $i$ of a network is denoted by $PR_i$ and it satisfies
\begin{equation}
PR_i=\frac{1-d}{N}+d\sum_{j=1}^N \frac{A_{ij}}{k_j}PR_j,
\end{equation} 
where $k_j$ is the degree of the node $j$ and $d$ is a tuning parameter usually set to $0.85$. The parameter $d$ interpolates among an uniform assignment ($d=0$) to a spectral centrality measure based on the Laplacian of the network $(d=1)$, defined as $L=AD^{-1}$, and describing the equilibrium measure of a random walk on the graph.

\subsection{Localization and IPR}

Detecting the eigenvectors of interest, typically the ones related to the eigenvalues at the edges of the spectrum, is the main problem in using spectral methods.
As shown in \cite{newman1}, spectral algorithms work well if the graph is sufficiently dense. In this case, $A$'s spectrum has a discrete part and a continuous part in the limit $N\to\infty$. Its first eigenvector essentially sorts vertices according to their degree, whereas the second eigenvector is correlated with the communities. 
The problem is when this second eigenvalue gets lost in the continuous bulk of eigenvalues coming from the randomness in the graph. This is what happens in  the sparse case, when the largest eigenvalues of $A$
are controlled by the vertices of highest degree, and the corresponding eigenvectors are localized around these vertices \cite{sudakov}. As $N$
grows, these eigenvalues swamp the community-correlated eigenvectors, if any, with the bulk of uninformative eigenvectors. As
a result, spectral algorithms based on $A$
fail at a significant distance
from the detectability threshold \cite{redemption}.

In core-periphery detection, this problem is not present, because the attention is focused on the eigenvector related to the first eigenvalue, which remains far from the bulk. For example, in a Erd\'os-Renyi graph it grows as the square root of the highest degree in the graph, and in the sparse case as $\log N/\log(\log N)$ \cite{sudakov}.  Nevertheless, localization of the eigenvector might be still a problem for some methods, such as those using a threshold (See \cite{lip}).  
A common measure of localization of the normalized $j$-th eigenvector, $\boldsymbol{v^j}$, is the inverse participation ratio (IPR), defined as
\begin{equation}\label{IPRdef} 
I_j= \sum_{i=1}^N (v^j_i)^4
\end{equation} 
If the components of an eigenvector are identical, $v^j_i=1/\sqrt{N}$ for every $i$, then $I_j=1/N$. For an eigenvector with only one non vanishing component, $v^j_i=\delta_{i,i'}$, the inverse participation ratio is $1$. The comparison of
these two extremal cases illustrates that with the help of the IPR, one can tell whether only
$\mathcal{O}(1)$ or as many as $\mathcal{O}(N)$ components of an eigenvector differ significantly from $0$, i.e., whether an eigenvector is localized or not.

In the  $c$-random regular graph the first eigenvalue is $c$ and the corresponding eigenvector is constant, $v^1_i=1/\sqrt{N}$, hence non-localized according to IPR. The same happens in the rSBM where the number of non zero components of the first eigenvector is $\mathcal{O}(N)$. In fact the first eigenvector of the adjacency matrix can be written as $v^1_i\propto u_{g(i)}$, being  $\boldsymbol{u_g}$ the first eigenvector of the affinity matrix $p_{ab}$, and the number of significantly non zero entries is at least the expected size of any community $\bar{g}$,  $\lambda_{\bar{g}}N$, for which $u_{\bar{g}}\neq 0$.

Slight modifications with respect to the fixed degree cases have to be considered in the standard Erdos-Renyi and SBM random graphs. In fact, as soon as the expected degree $c\gg 1$ (at least $c=\mathcal{O}(\log(N))$ \cite{erdos1}), all the eigenvectors are shown to be delocalized, being all the components bounded by an $\mathcal{O}(1/\sqrt{N})$ term. In particular the first eigenvector of the adjacency matrix differs from the correspondent at fixed degree (rr and rSBM) for a random Gaussian noise of width $\mathcal{O}(1/\sqrt{N})$. In the completely sparse case, $c=\mathcal{O}(1)$, at best of our knowledge, there are no rigorous results concerning localization of the eigenvectors. However in \cite{barabasi1} it is shown how the first eigenvector of the uncorrelated sparse random graph is still IPR-delocalized as opposed to the other eigenvectors that are much more localized, due to the presence of isolated nodes in the graph.

Delocalization in the first eigenvector of the SBM adjacency matrix (and the other spectral centrality measures) explains why  MINRES and  eigenvector centrality tend to be equivalent in the limit $N\to \infty$. The situation critically changes in sparse graphs with non Poissonian degree distribution, more similar to real networks. For example the eigenvectors related to a
scale-free graph's largest and smallest eigenvalues are highly localized on the vertices with highest degree \cite{barabasi1}. This transition to localization was well-explained in Ref. \cite{newman2} with a toy model composed of a random graph plus a single hub node: when the degree of the hub becomes bigger than the Poissonian mean square of the one in the rest of the graph, the first eigenvector starts to localize on the hub node and their neighbors. In Section \ref{sec:result} we generate core periphery networks with heterogeneous core using dcSBM, testing the correlation between the spectral centrality measures performances, measured in terms of overlap with the original assignment, and  their IPR as index of localization.

\section{Results} \label{sec:result}
\subsection{SBM core-periphery networks}\label{sec:hom}
\begin{figure}
\centering
\includegraphics[width=90mm]{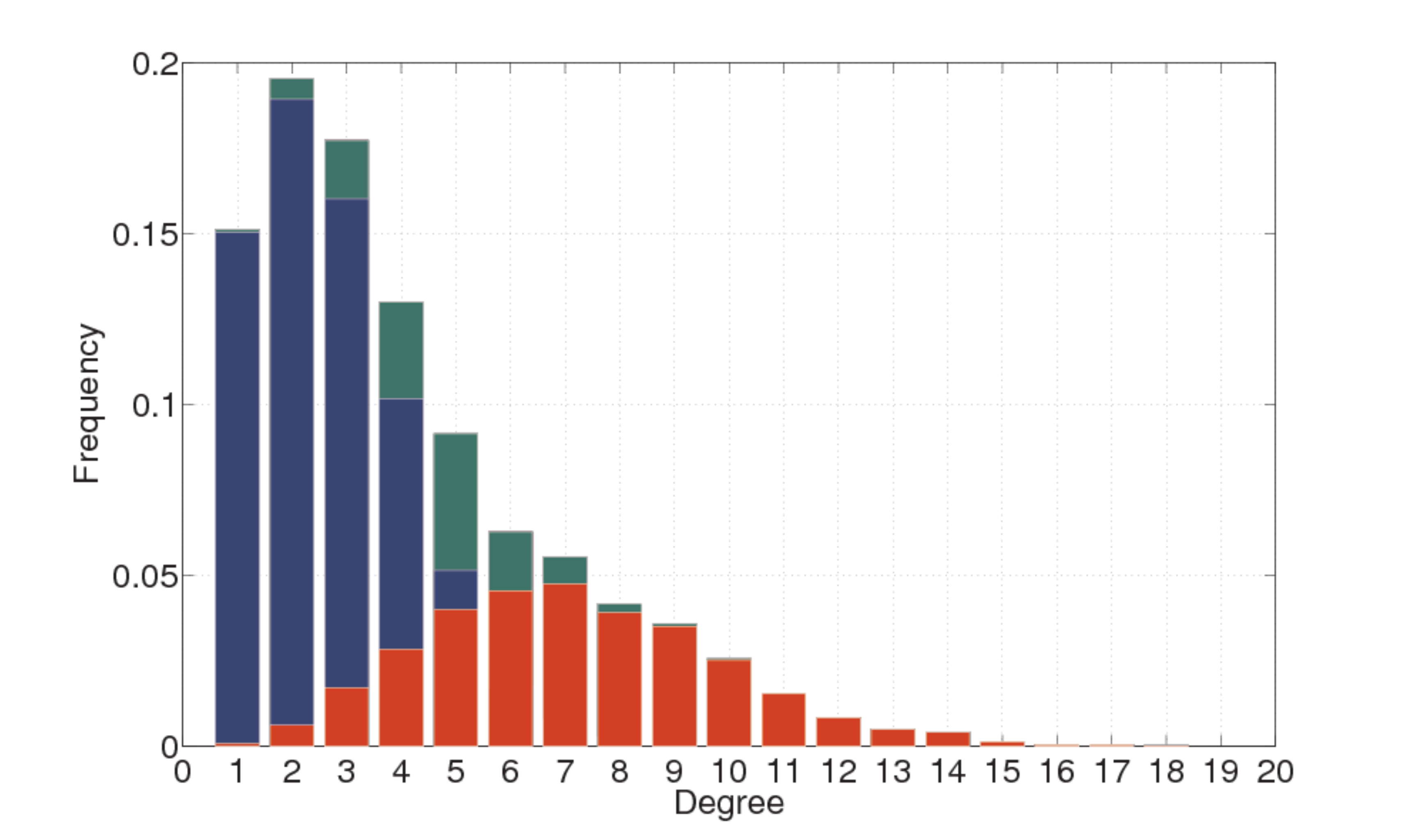}
\protect\caption{Degree-distributions of core (red) and periphery (blue) nodes in a core-periphery SBM network of size $N = 10^4$, core-fraction $\gamma=0.3$ and affinity matrix $c_{ab} =[10\,\, 6; 6\,\,1]$.}
\label{FIGDEG}
\end{figure}

We investigate networks sampled from a SBM with a specific core-periphery structure where the elements of the affinity matrix satisfy
$$\frac{c_{11}}{c_{12}} \neq \frac{c_{12}}{c_{22}},$$
so that  BP marginal centrality differs from pure degree centrality (see equation ($\ref{eq:bp-update-1}$)).
Even if the expected difference between the degree of a core-node and of a periphery-node is large, since in the $N\to \infty$ limit the degree distribution is a mixed-Poissonian, we still expect   a finite fraction of core-nodes to share the same degree with a finite fraction of periphery-nodes and we expect a finite difference between the ranking error given by degree-centrality and by the probability marginals. We show the results for a core-periphery SBM network with core-fraction $\gamma=0.3$ and affinity matrix $c_{ab} =[10\,\, 6; 6\,\,1]$. Figure \ref{FIGDEG} shows the degree distribution for the case with $N=10^4$ nodes, indicating a significant overlap of the degree of core- and periphery-nodes. 

\begin{figure}
\begin{tabular}{cc}
  \includegraphics[width=75mm]{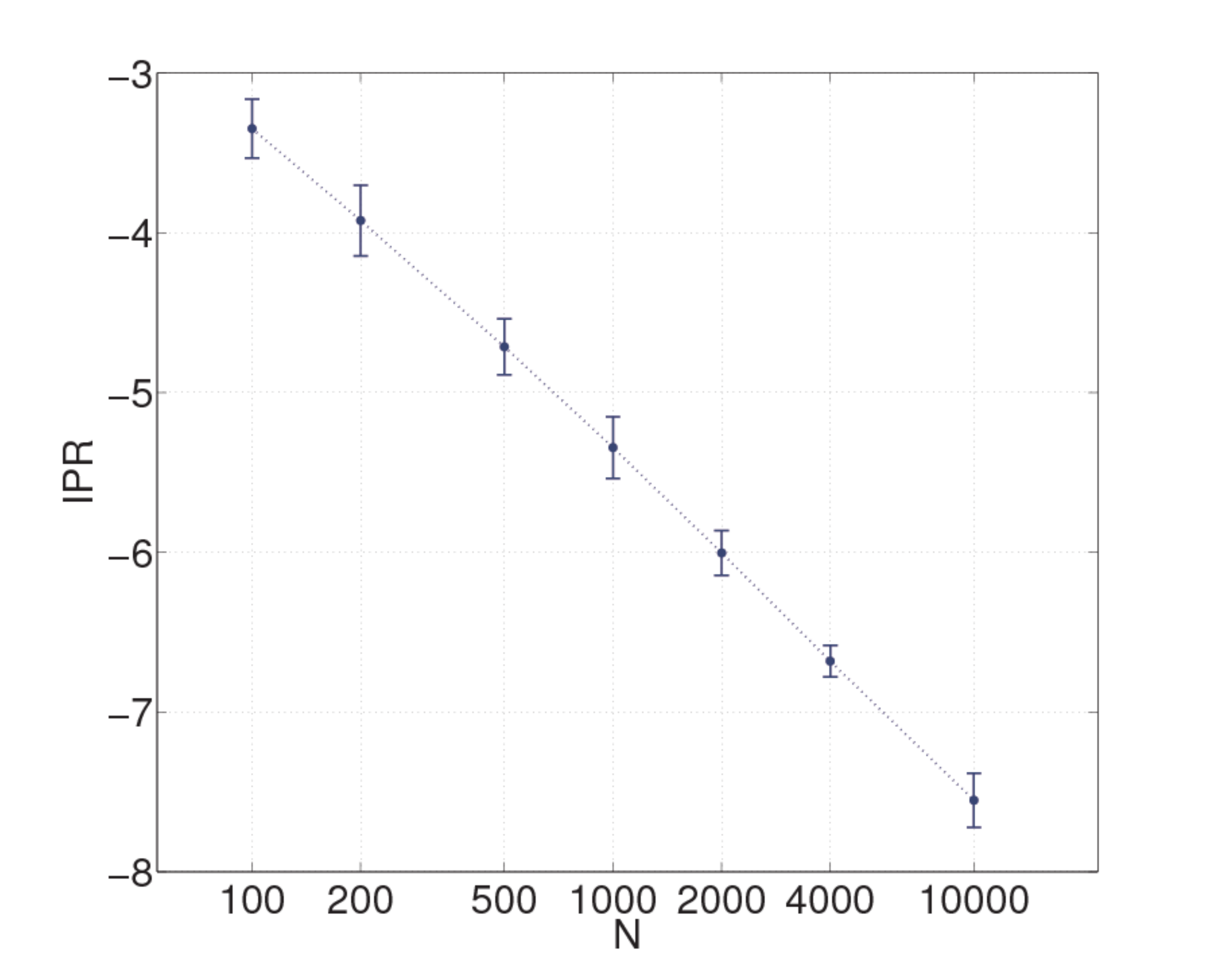} &   \includegraphics[width=75mm]{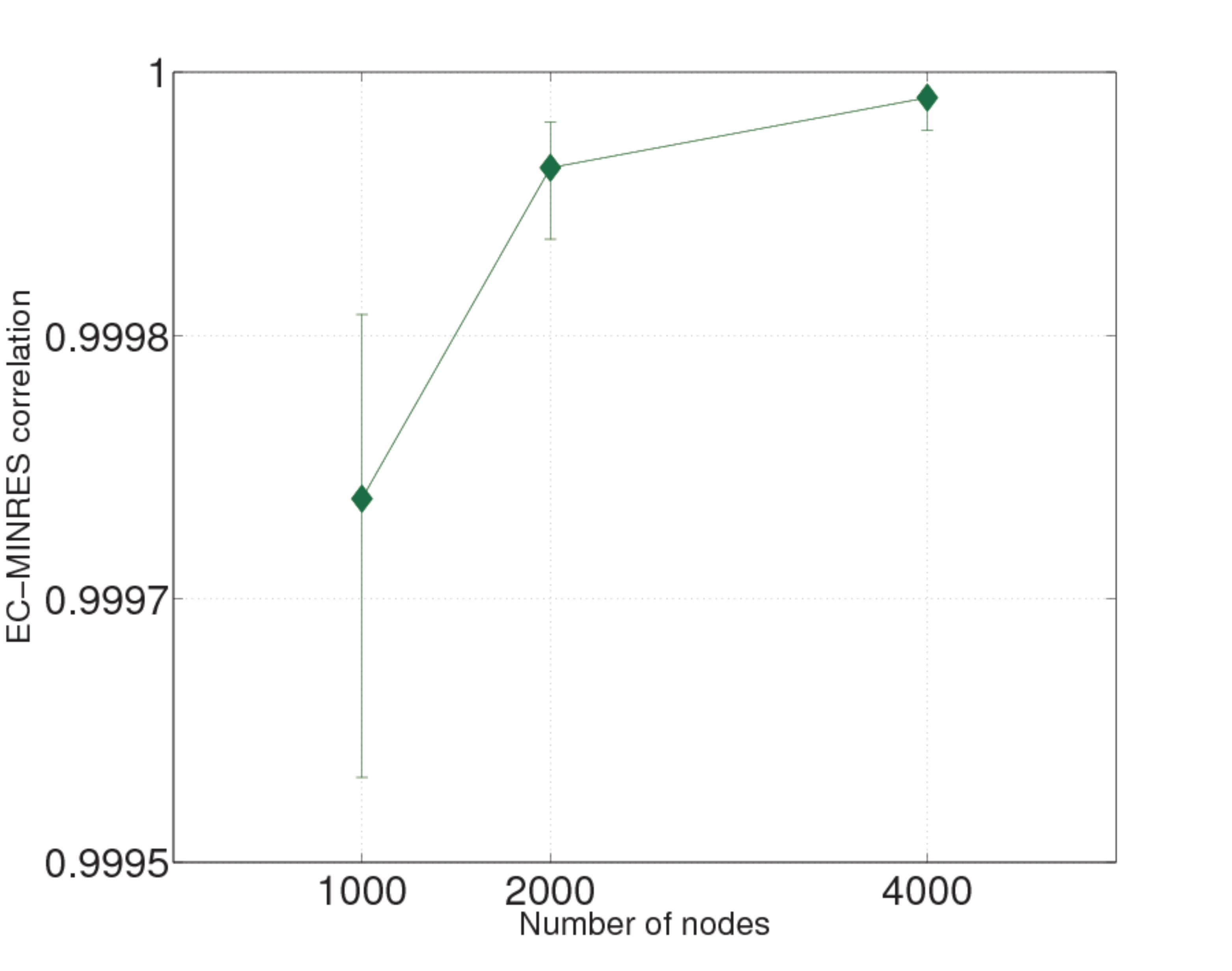} \\
(a) & (b)\\[6pt]
\end{tabular}
\caption{(a) Size-scaling of the IPR of Eigenvector centrality (EC). Simulations on 100 core-periphery SBM networks with core-fraction $\gamma=0.3$ and affinity matrix $c_{ab} = [10\,\, 6; 6\,\,1]$. (b) Pearson correlation between EC and MINRES  coreness with the same parameters. Data-points correspond to the medians, while error bars are interquartile ranges. }\label{fig:ECMIN}
\end{figure}

As a preliminary analysis, we investigate the relation between eigenvector centrality and MINRES. Left panel of Figure \ref{fig:ECMIN} shows that the IPR of EC of a core periphery SBM tends to zero in the large $N$ limit. Following the argument above (see eq. ($\ref{derBOYD-1}$)), we expect that for this model MINRES centrality converges to EC in the same limit. The right panel of Figure \ref{fig:ECMIN} shows that indeed the Pearson correlation between EC and MINRES tends to one for large network size. Therefore for large SBM the MINRES centrality is equivalent to EC. For this reason we will not consider the former. 

For each centrality/coreness measure $\mathcal{M}$ we consider the corresponding score vector $\bold{v^\mathcal{M}}$, we sort its values and evaluate $C^{\mathcal{M}}$, i.e. the set of nodes that are most likely to be in the core according to $\mathcal{M}$. These are simply  the nodes  corresponding to the $\gamma N$ largest entries in $\bold{v^\mathcal{M}}$.
We define the assignment vector $\bold{t^\mathcal{M}}$, that equals one for components inside $C^\mathcal{M}$ and zero otherwise and we measure the performance of the measure $\mathcal{M}$ by computing the \textit{agreement} between $\boldsymbol{t^\mathcal{M}}$ and the original assignment $\boldsymbol{g}$, i.e. $A(\bold{\boldsymbol{g},t^\mathcal{M}})=1/N\sum_{i=1}^N \delta_{t^\mathcal{M}_i,g_i}$.
We also introduce a normalized agreement that we call overlap, defined as
\begin{equation}
q(\bold{\boldsymbol{g},t^\mathcal{M}}) =  \frac{A(\bold{\boldsymbol{g},t^\mathcal{M}})- \max_a \gamma_a}{1 - \max_a \gamma_a}.
\end{equation}
The overlap is defined so that it is equal to $1$ when the  labeling is correct, and zero if the only information we have are the group sizes $\gamma_a$ and we assign each node to the largest group to maximize the probability of getting the correct assignment.

\begin{figure}[h!]
\centering
\includegraphics[width=90mm]{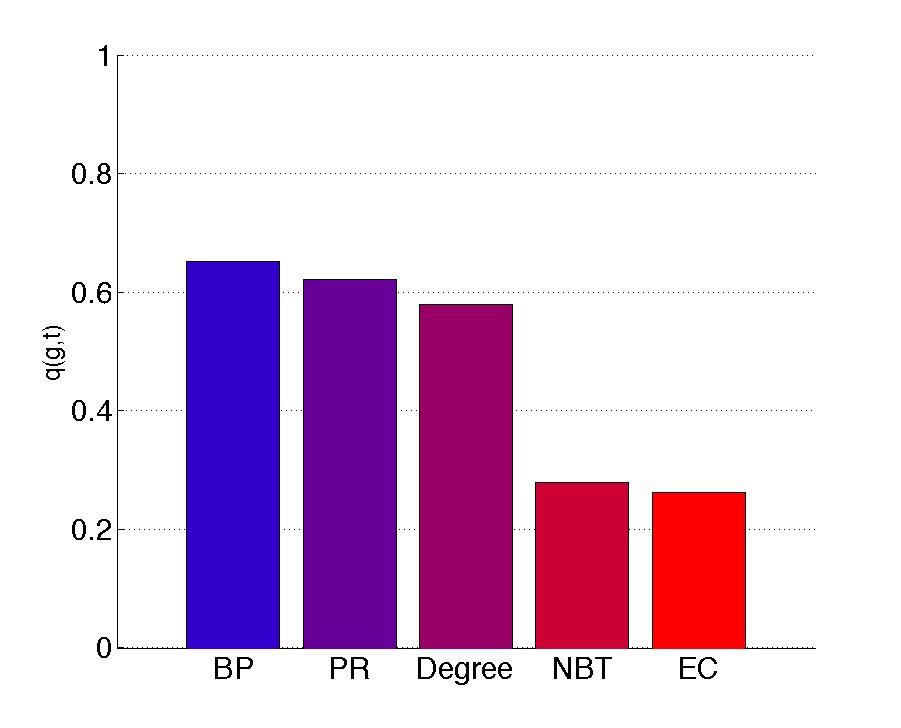}
\protect\caption{Ranking of the methods in the homogeneous case. Simulations on 100 core-periphery SBM networks of size $N = 10^4$, core-fraction $\gamma=0.3$ and affinity matrix $c_{ab} = [10\,\, 6; 6\,\,1]$}\label{fig:RANK}
\end{figure}

 Fig.\ref{fig:RANK} shows the overlap for the five metrics. As expected, BP gives the best assignment, since it is based on the maximization of the likelihood of the generative method. Quite surprisingly PageRank and degree centrality performs only slightly worse than BP, thus these metrics give a good indication of coreness. NBT and EC perform significantly worse, both having an average overlap below $0.8$.    
 
\begin{figure}[h!]
\centering
\includegraphics[width=60mm]{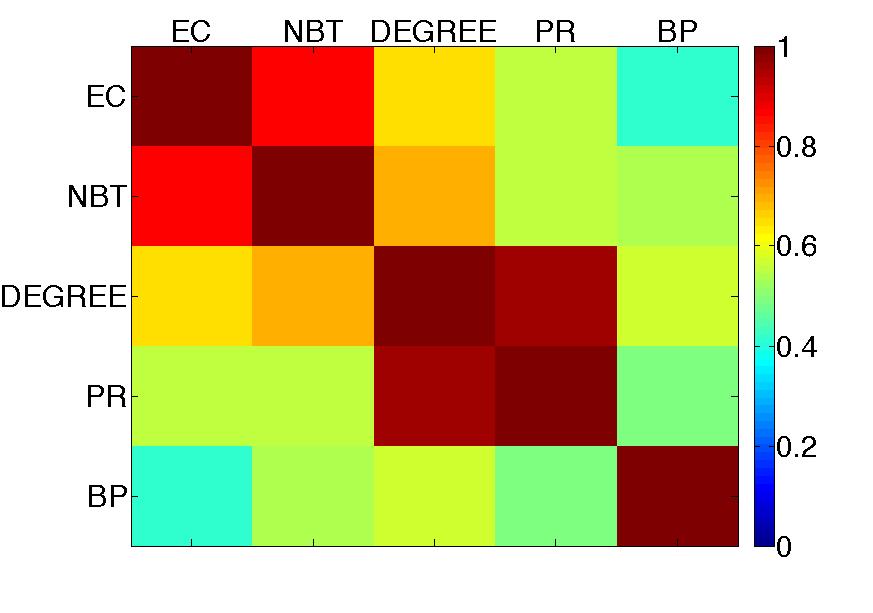}
\includegraphics[width=60mm]{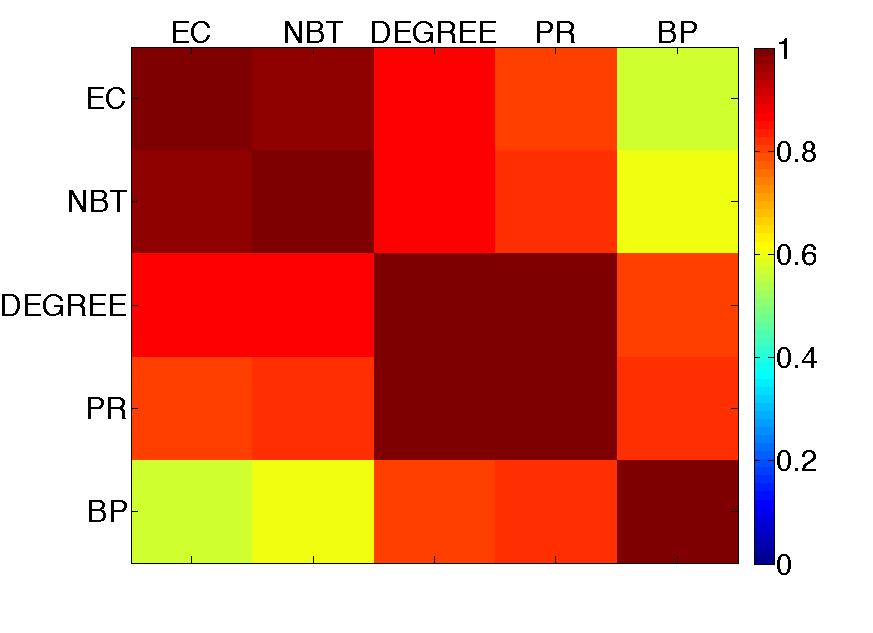}
\includegraphics[width=60mm]{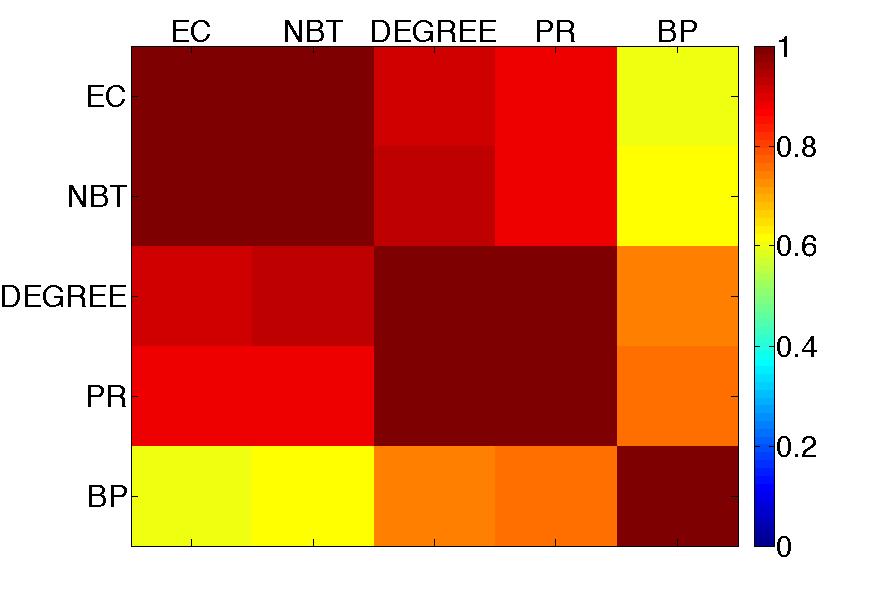}
\protect\caption{Matrix of Pearson cross-correlations between the investigated measures. (EC) Eigenvector Centrality, (NBT) Non-backtracking Centrality, (DC) Degree Centrality, (PR) PageRank, (SBM) SBM marginals.
Simulations on 100 core-periphery SBM networks of size $N = 10^4$,   affinity matrix $c_{ab} = [10\,\, 6; 6\,\,1]$ and different size core-fraction $\gamma=0.1,0.3,0.5,0.7$ (top-left, top-right, bottom-left, bottom-right).}\label{fig:PEARSON}
\end{figure}

In order to better characterize the mutual relations between two metrics, we introduce the Pearson correlation of the corresponding centralities, measuring the correlation of the assigned scores to a node. We also investigate a range of core size, ranging from $\gamma=0.1$ to $\gamma=0.7$. Through these correlations, see Fig.\ref{fig:PEARSON}, we confirm the similarity between Eigenvector Centrality and Non-backtracking centrality and the strong correlation between Degree centrality and PageRank, despite the assignment errors of the latter are slightly lower. We find that BP marginals differ from all other methods. This difference can be understood considering the low IPR related to this measure. In fact marginals tend to be strongly polarized to zero or one within the two groups thus all finer correlations with the other methods related to properties different from group assignment are lost.  

\subsection{Degree corrected SBM core-periphery networks}\label{sec:het}
In this section we investigate the role of degree heterogeneity in the measurement of centrality properties. In fact, as explained above, SBM has a mild degree heterogeneity, while many real networks display fat tailed degree distribution, To this end we sample networks from a degree corrected SBM with a strong core-periphery structure but an heterogeneous core having a power law degree distribution. This is done also with the idea that being member of a core in a complex network creates a positive-feedback loop that enhances the probability of being a hub for core-members and to avoid hubs in the periphery.

\begin{figure}
\begin{tabular}{cc}
  \includegraphics[width=75mm]{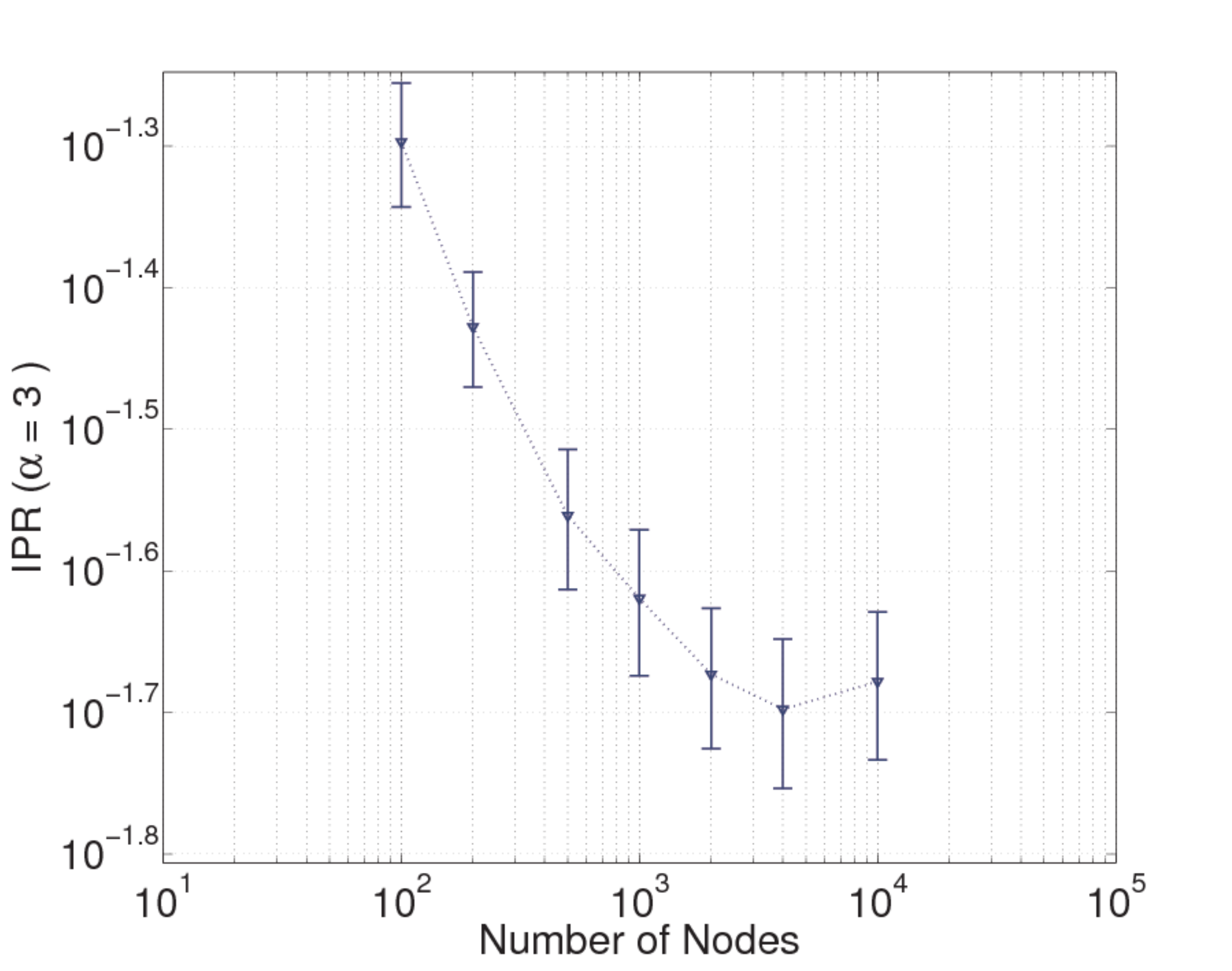}   & \includegraphics[width=75mm]{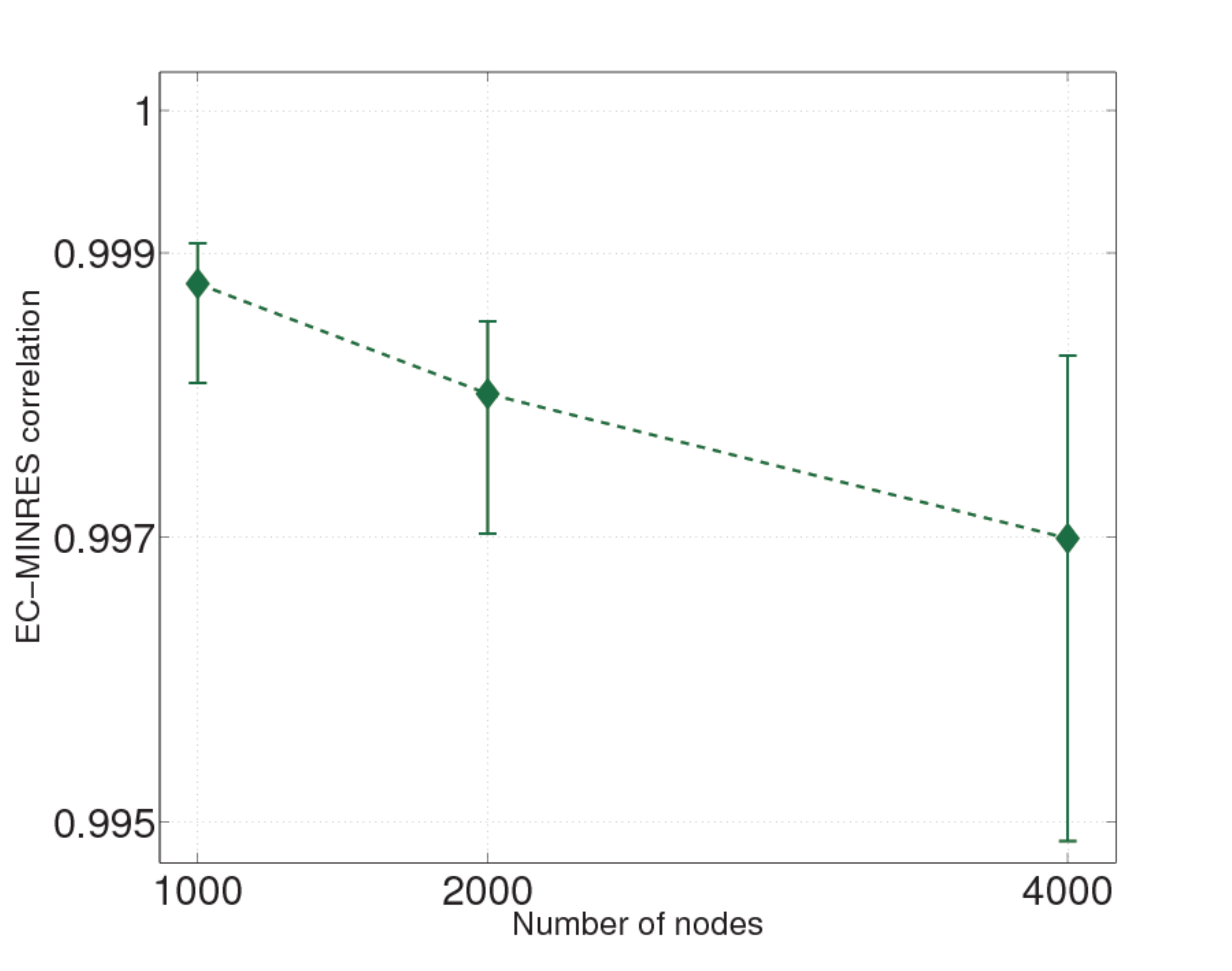} \\
(a) & (b)\\[6pt]
\end{tabular}
\caption{(a) Size-scaling of the IPR of Eigenvector centrality. Simulations on 100 core-periphery dcSBM networks with core-fraction $\gamma=0.3$, affinity matrix $c_{ab} = [10\,\, 6; 6\,\,1]$, and $\alpha = 3.0$. (b) Pearson correlation between Eigenvector centrality and MINRES coreness. Simulations on 20 core-periphery dcSBM networks with the same parameters. Data-points correspond to the medians, while error bars are interquartile ranges.}\label{fig:ECMIN2}
\end{figure}

In a power law degree distributed network the first eigenvector of the adjacency matrix is localized as soon as the typical eigenvalue given by the average behavior of the network $\bar{\lambda}\sim \langle k^2\rangle/\langle k\rangle\sim N^{(3-\alpha)/(\alpha-1)}$ is overcome by the one relative to the highest degree $\lambda_M\sim \sqrt{k_M}\sim N^{1/2(\alpha -1)}$, thus for $\alpha > 2.5$. In this regime the inverse participation ratio of the spectral centrality measures is definitely larger than zero, while it tends to zero in the limit $\alpha\to \infty$, where one recovers the homogeneous case. 
For what we have seen in Section \ref{sec:method}, in this regime EC and MINRES are expected to give different centrality measures. Fig. \ref{fig:ECMIN2} confirms this intuition because when $N\to \infty$ IPR of EC does not converge to zero and the EC-MINRES correlation decreases. Moreover we found that MINRES centrality vector is always more localized and its overlap with the original assignment is always smaller than EC's overlap.

\begin{figure}
\begin{tabular}{cc}
    \includegraphics[width=75mm]{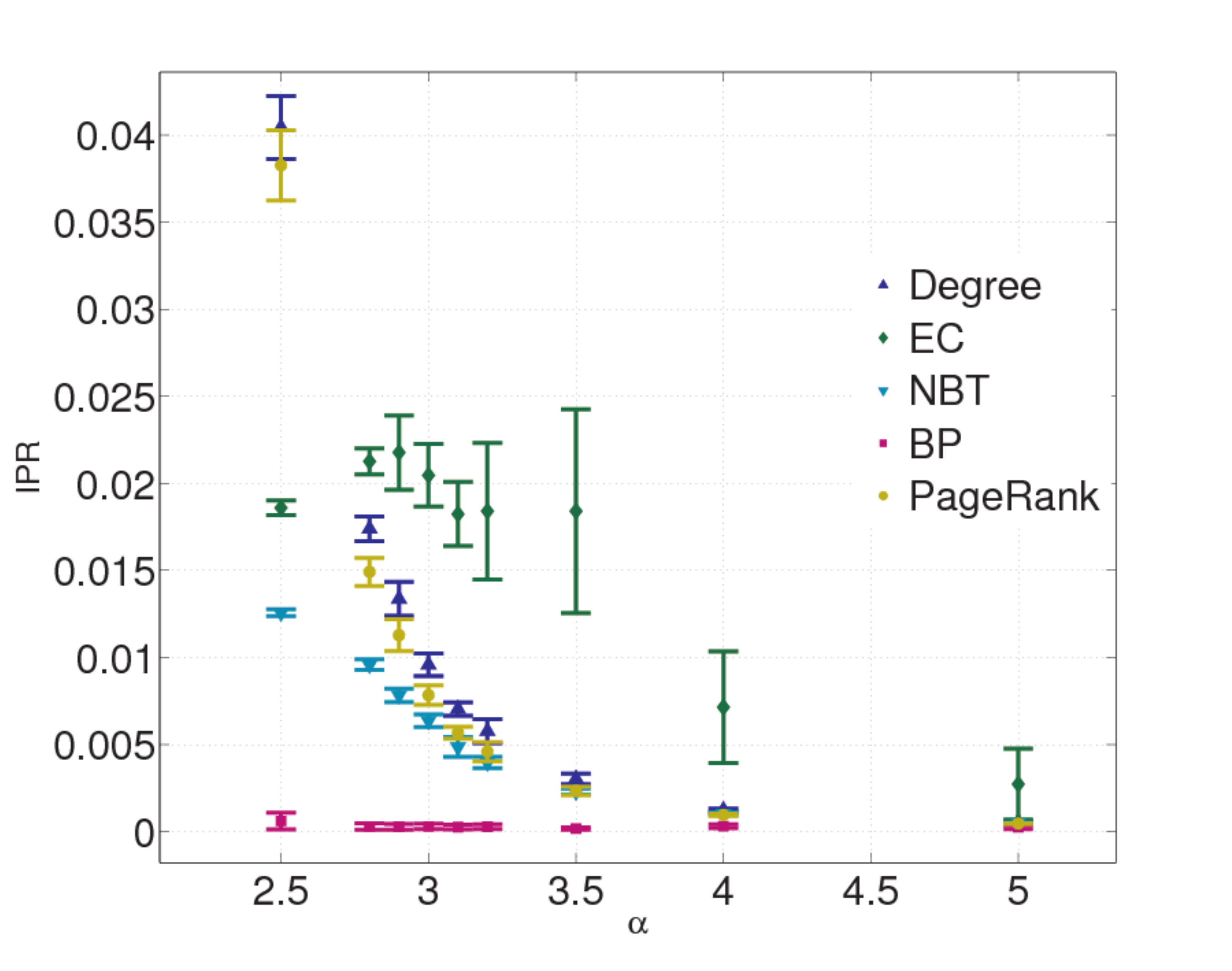} &  \includegraphics[width=75mm]{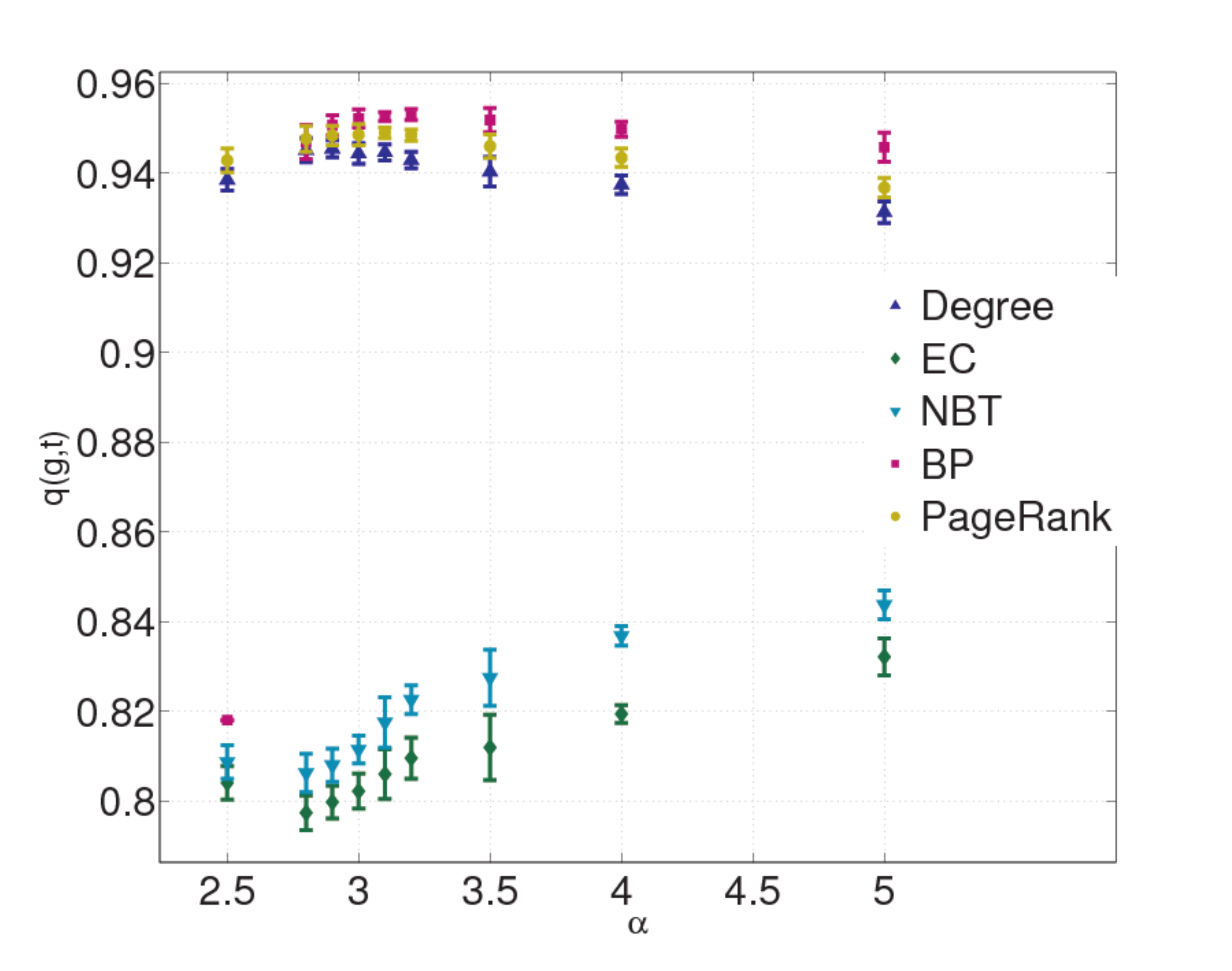} \\
(a)& (b)\\[6pt]
\end{tabular}
\caption{(a)  IPRs of the centrality measures (rescaled so that  $\sum_{i=1}^N (v_i)^2 = 1$) as a function of the heterogeneity parameter $\alpha$. Simulations on 100 core-periphery SBM networks
of size $N = 10^4$, core-fraction $\gamma=0.3$, and affinity matrix $c_{ab} = [10\,\, 6; 6\,\,1]$.  (b) Overlap of different centrality measures with the original assignment as a function of the heterogeneity $\alpha$. Simulations on 100 core-periphery SBM networks
with the same parameters as before.}\label{fig:IPRs}
\end{figure}

The left panel of Figure \ref{fig:IPRs} shows IPR for the different centrality metrics as a function of the tail exponent $\alpha$. As expected the localization is higher when $\alpha$ is small. Moreover NBT has the smallest IPR because the absence of reflection mechanism alleviates the problem of localization.  
Another possible way to work around the localization problem in networks has been found in \cite{PR1} by adding a regularizing term to every adjacency matrix element, as if there were a weak edge between every pair of nodes. The effect of this regularization has been proven to be equivalent to PageRank \cite{PR2}. Our analysis shows that indeed PageRank localization is, among  all the centrality measures,  just higher then NBT localization and its performance is comparable with the ones by BP marginal and degree centrality.

The right panel of Figure  \ref{fig:IPRs} compares the performance  of the different centrality measures in identifying the core nodes. As for the homogenous case, BP is the best method, but Degree and PageRank have comparable performance. Despite the fact SBM is not the generative methods, BP based on it recovers well the core nodes, at least when degree heterogeneity s not extreme. Finally, NBT and EC perform significantly worse.

\begin{figure}[h!]
\begin{tabular}{cc}
  \includegraphics[width=70mm]{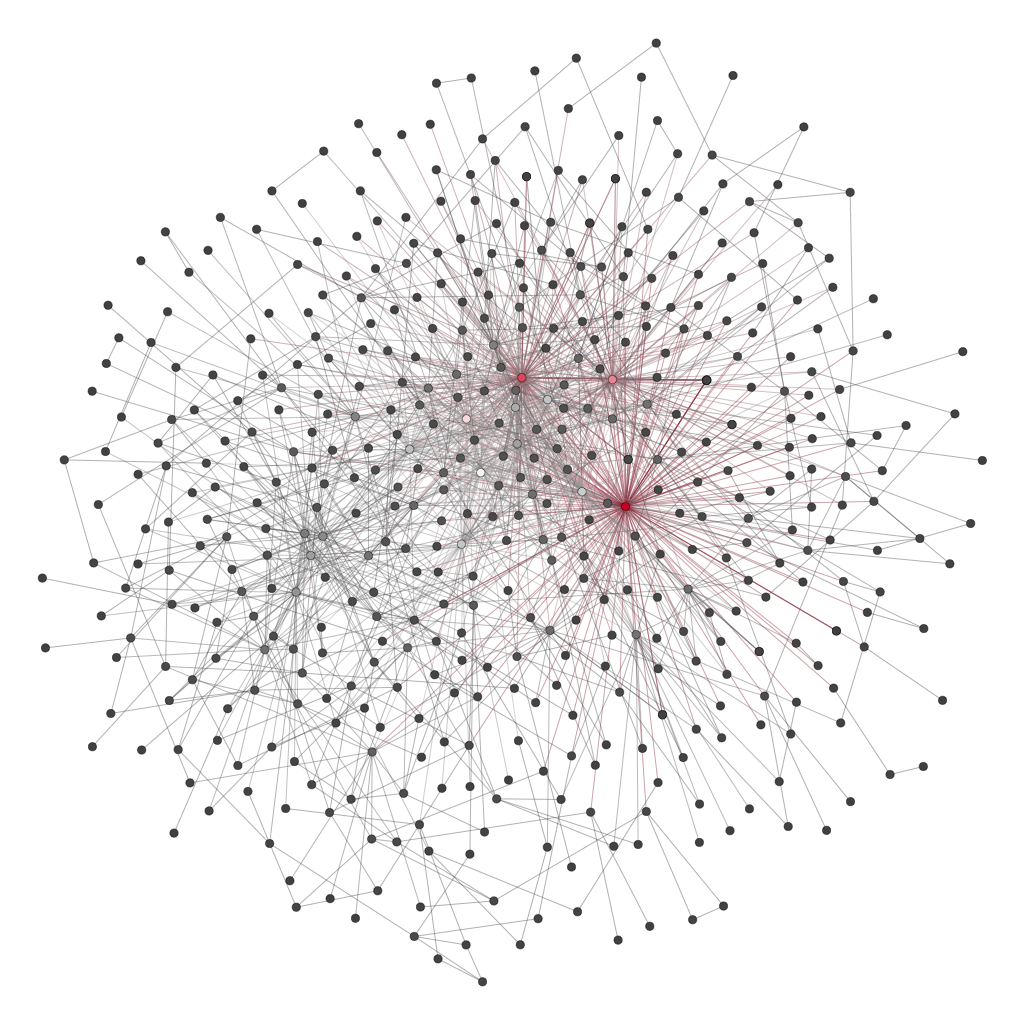} &   \includegraphics[width=90mm]{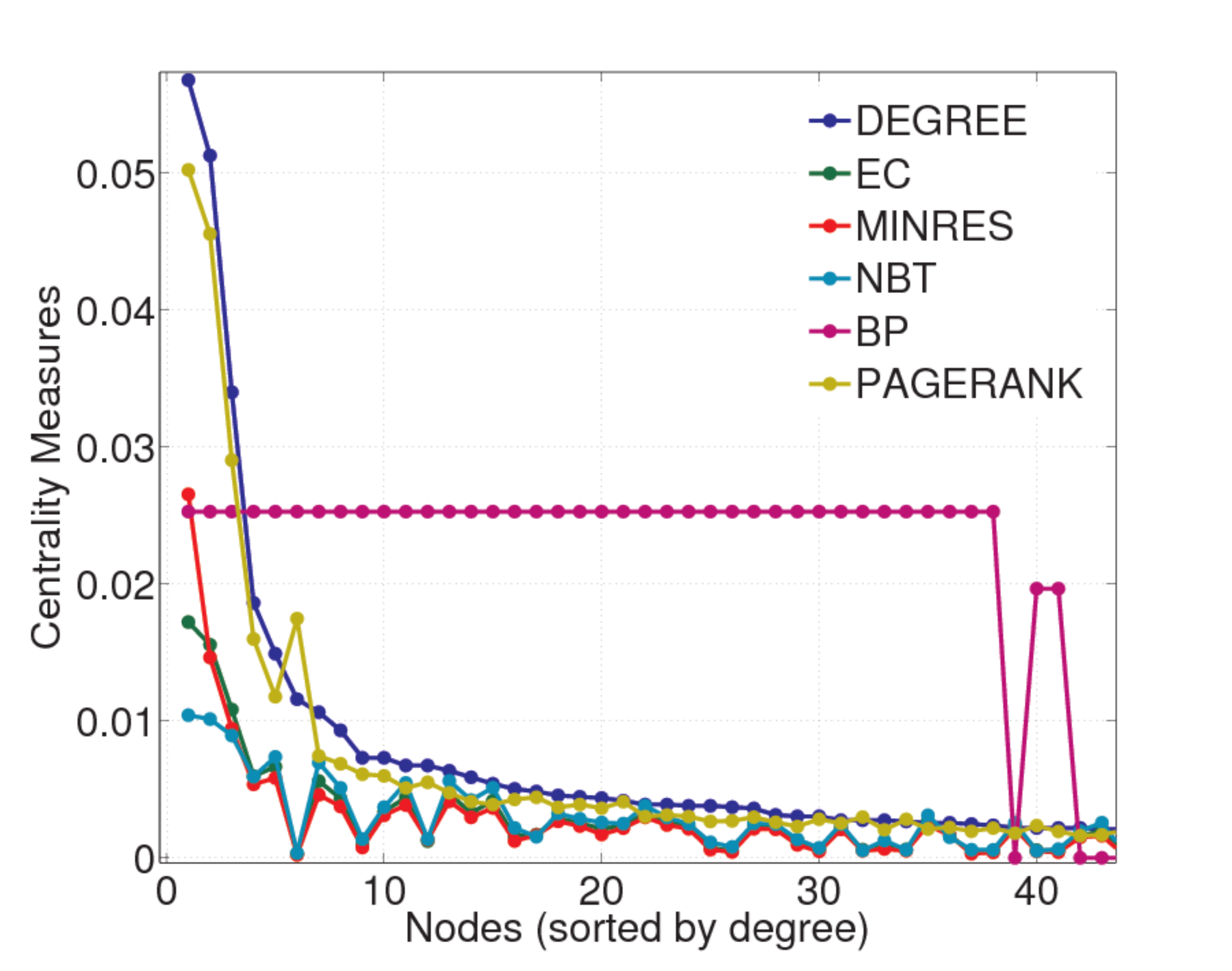} \\
(a) & (b) \\[6pt]
\end{tabular}
\caption{(a) Graphic representation of the 1000 largest-degree nodes subnetwork of the Internet (snapshot of  the 20-11-1997 with 3066 nodes) at the level of autonomous systems from Oregon Route Project\cite{oregon}.  Nodes are colored according to the degree (large-degree nodes are red). (b) Values of the centrality measures over different nodes, sorted by decreasing degree. 
}\label{fig:OREGON}
\end{figure}

\begin{figure}[h!]
\centering
\includegraphics[width=90mm]{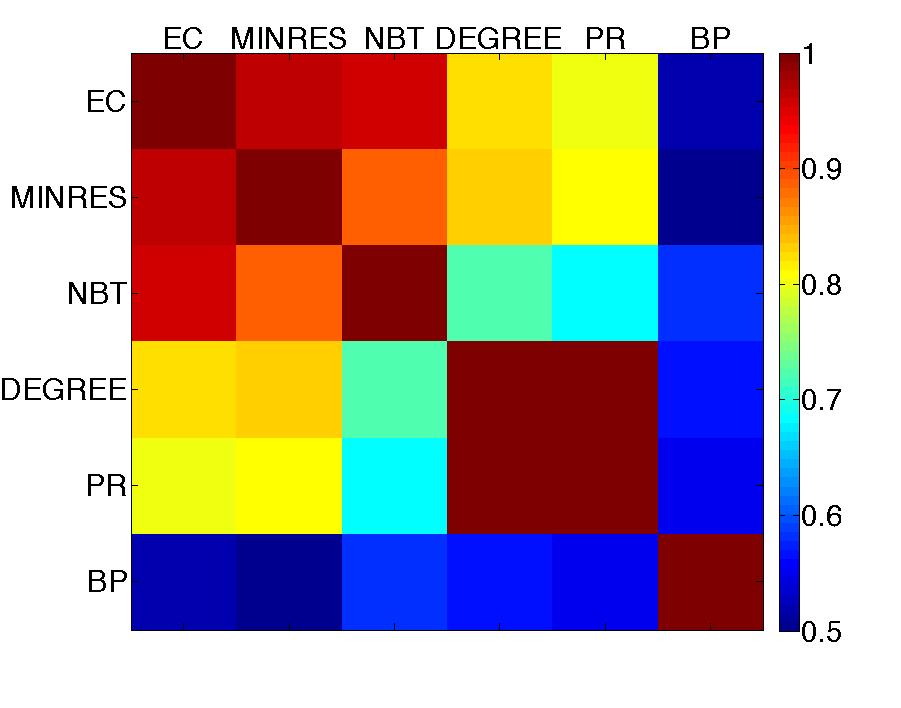}
\protect\caption{Matrix of Pearson cross-correlations between the centrality measures on the Autonomous Systems network (Oregon Route Project) . (EC) Eigenvector Centrality, (MINRES) Minres coreness measure, (NBT) Non-backtracking Centrality, (DC) Degree Centrality, (PR) PageRank, (BP) BP marginals. 
}\label{fig:PEAROREGON}
\end{figure}

To show the robustness of results on the properties of the centrality measures and on their mutual correlations we analyze one snapshot of the network constructed from data of the Oregon Routeviews Project (see left panel of Fig.\ref{fig:OREGON}) representing the Internet at the level of Autonomous Systems. 
This network has already showed to have clear core-periphery structure \cite{key-35}: it is generally composed  of a large number of leaves, typically
client autonomous systems corresponding to end users like ISPs, corporations, or educational institutions,  and a small number of well-connected nodes \cite{real1,real2}. 
The right panel of Fig.\ref{fig:OREGON} shows the centrality measure of the nodes sorted by degree along the x axis. We cannot compute the overlap with a real assignment but the comparison of the different methods gives us a precise and significant indication on the set of nodes to be considered part of the core in this network. BP identifies a core of roughly 40 nodes, while the other measures are, as in numerical simulations, clustered in two groups,  Degree and PageRank in one and EC, MINRES, and NBT in the other.  Finally Fig.\ref{fig:PEAROREGON} shows the Pearson correlation matrix of the six methods. This figure confirms (i) the clustering in two groups, (ii) the  strong similarity between Degree and PageRank, and the fact that BP is fairly uncorrelated with the other methods.

\section{Conclusions} \label{sec:concl} 
In this work we have connected centrality and coreness measures and investigated the differences among them on a set of core-periphery graphs and on a \textit{ad-hoc} set of graphs designed to be both core-periphery structured and degree-heterogeneous. In these models the affinity matrix has a clear core-periphery structure and high degree-heterogeneities were assigned with a strong positive correlation with core-membership. 
We have characterized centrality measures both in terms of their localization properties, quantified by the inverse participation ratio, and in terms of their ranking error of the assignment to the core at increasing levels of degree-heterogeneity. 

In the homogeneous case we verified that SBM probability marginals obtained with belief propagation outperform all other measures in predicting the core-members, also being the only method able to estimate the size of the core non-parametrically. This is expected since the marginals are obtained by maximizing the likelihood of the generative model. Interestingly PageRank and degree centrality perform very well in identifying core nodes, while eigenvector centrality and non backtracking centrality perform worse. 
The results on the ranking of centrality measures are quite robust to the presence of heterogeneity in degree as modeled by a degree corrected SBM. Despite SBM is not anymore the generative model, its marginals (obtained with BP) give the best assignment and slightly outperforms PageRank and degree centrality. Remarkably, marginals of SBM perform very poorly when the degree heterogeneity is very large (i.e. when the tail index of the degree distribution is below 3). PageRank and degree centrality are the best methods in this regime. Finally, we have investigated the relation between Eigenvector Centrality and MINRES. For the case of homogeneous graphs we have also observed the convergence in the large-networks limit these two methods caused by the vanishing inverse participation ratio of Eigenvector Centrality in SBM graphs.
In the heterogeneous case we have found no general relation between localization and ranking error although an anti-correlation holds if we restrict to Eigenvector centrality, Non-backtracking centrality and MINRES coreness.

In this work we have focused on centrality measures in graphs with a core-periphery structure and varying degree-heterogeneity. In the last few years many other local properties of networks have been introduced and used in applications but their expected values and mutual relations in benchmark graph-ensembles are still unknown, thus further analytical and numerical research should be addressed in quantifying relations between different graph properties in realistic directed, group-structured and heterogenous graph-ensembles. In particular rigorous results are needed on the localization and assignment properties of different spectral centrality measures and their relation with the behavior of the eigenvalue spectrum.

\section*{Acknowledgment}

 PB and FL acknowledge partial support by the grant SNS13LILLB ''Systemic risk in financial markets across time scales". This work is supported by the European Community's H2020 Program under the scheme "INFRAIA-1-2014-2015: Research Infrastructures", grant agreement \#654024 "SoBigData: Social Mining \& Big Data Ecosystem'"(http://www.sobigdata.eu). Authors also thanks Travis Martin for useful discussions.

\end{document}